\documentclass{ws-ijmpcs}
\usepackage{slashed}
\begin{document}

\markboth{XIU-LEI REN et al.}
{LOWEST-LYING OCTET BARYON MASSES ......}

%%%%%%%%%%%%%%%%%%%%% Publisher's Area please ignore %%%%%%%%%%%%%%%
%
\catchline{}{}{}{}{}
%
%%%%%%%%%%%%%%%%%%%%%%%%%%%%%%%%%%%%%%%%%%%%%%%%%%%%%%%%%%%%%%%%%%%%

\title{LOWEST-LYING OCTET BARYON MASSES
IN COVARIANT BARYON CHIRAL PERTURBATION THEORY}

\author{XIU-LEI REN}

\address{School of Physics and
Nuclear Energy Engineering, Beihang University, Beijing, 100191 China\\
xiuleiren@phys.buaa.edu.cn}

\author{LISHENG GENG}

\address{School of Physics and
Nuclear Energy Engineering, Beihang University, Beijing, 100191, China\\
International Research Center for Nuclei and Particles in the Cosmos, Beihang University, Beijing, 100191, China\\
lisheng.geng@buaa.edu.cn}

\author{JIE MENG}

\address{School of Physics and
Nuclear Energy Engineering, Beihang University, Beijing, 100191, China\\
State Key Laboratory of Nuclear Physics and Technology, School of Physics, Peking University, Beijing, 100871, China\\
Department of Physics, University of Stellenbosch, Stellenbosch, 7602, South Africa\\
mengj@pku.edu.cn}

\author{HIROSHI TOKI}
\address{Research Center for Nuclear Physics (RCNP), Osaka University, Ibaraki, Osaka, 567-0047, Japan\\
toki@rcnp.osaka-u.ac.jp}

\maketitle

\begin{history}
\received{Day Month Year}
\revised{Day Month Year}
\end{history}

\begin{abstract}
We report on a systematic study of the ground-state octet baryon masses in the covariant baryon chiral perturbation theory with the extended-on-mass-shell renormalization scheme up to next-to-next-to-next-to-leading order,
taking into account the contributions of the virtual decuplet baryons.
A reasonable description of the lattice results is achieved by fitting  simultaneously all the publicly available $n_f = 2+1$ lattice QCD data. It confirms that the various lattice simulations are consistent with each other. We stress that
a self-consistent treatment of finite-volume corrections is important to obtain a $\chi^2/\mathrm{d.o.f.}$ about 1.

\end{abstract}

\ccode{PACS numbers: 12.39.Fe,  12.38.Gc, 14.20.Gk}

\section{Introduction}	
In recent years,  the lowest-lying octet baryon spectrum  has been studied on the lattice with $n_f=2+1$ setups
by a number of lattice QCD (LQCD) Collaborations~\cite{Durr:2008zz}\cdash\cite{Beane:2011pc}. At present, most LQCD calculations still need to adopt larger than physical light-quark masses and relatively small volumes.
Therefore, the obtained lattice results have to be extrapolated to the physical world by performing the so-called ``chiral extrapolation"~\cite{Leinweber:2003dg}\cdash\cite{Bernard:2005fy} and taking into account finite-volume corrections (FVCs)~\cite{Gasser:1986vb,Gasser:1987zq}.

Baryon chiral perturbation theory (BChPT), as an effective field theory, provides a useful framework to perform such extrapolations.
In the past decades, the ground-state (g.s.) octet baryon masses have been studied extensively~\cite{Jenkins:1991ts}\cdash\cite{Bruns:2012eh}. However, one found that in the SU(3) sector, the convergence of BChPT is slow.
On the other hand, up to now, a simultaneous description of LQCD data with finite-volume
effects taken into account self-consistently is still missing. Such a study is necessary for a clarification of the convergence problem and for testing the consistency between different LQCD simulations. Furthermore,
it helps to determine/constrain the many unknown low-energy constants (LECs) of the BChPT at next-to-next-to-next-to-leading order (N$^3$LO).

In this work we report on the first systematic study of  the g.s. octet baryon masses in the extended-on-mass-shell (EOMS) BChPT up to N$^3$LO, taking into account the contributions of the virtual decuplet baryons. FVCs to the lattice data are calculated self-consistently. In order to fix all the $19$ LECs and test the consistency of the current lattice calculations, we perform a simultaneous fit to all the publicly available $n_f=2+1$ LQCD data from the PACS-CS~\cite{Aoki:2008sm}, LHPC~\cite{WalkerLoud:2008bp}, HSC~\cite{Lin:2008pr}, QCDSF-UKQCD~\cite{Bietenholz:2011qq}, and NPLQCD~\cite{Beane:2011pc} Collaborations.
\section{Results and Discussions}

The details of the studies can be found in Refs.~\cite{Ren:2012aj,Ren:2013dzt}. Here we only briefly summarize the main results.
Up to N$^3$LO, there are $19$ LECs to be determined. In order to ensure that the N$^3$LO BChPT is suitable for the description of the LQCD data, we have chosen only the LQCD data satisfying $M^2_{\pi}< 0.25$ GeV$^2$ and $M_{\phi}L > 4$. By performing a $\chi^2$ fit to the LQCD data and the experimental octet baryon masses, we obtain the LECs and $\chi^2/\mathrm{d.o.f.}$ shown in Table~\ref{Tab:fitcoef}.

\begin{table}[t]
\centering
\tbl{Values of the LECs and the corresponding $\chi^2/\rm{d.o.f.}$ from the best fits.
 We have performed fits  to the LQCD and experimental data at $\mathcal{O}(p^2)$, $\mathcal{O}(p^3)$, and $\mathcal{O}(p^4)$ , without ($\slashed D$) and with ($D$) the explicit contributions of the virtual decuplet baryons.}
{\begin{tabular}{cccc|c}
\toprule
         & $\slashed D$-$\mathcal{O}(p^2)$ & $\slashed D$-$\mathcal{O}(p^3)$  &   $\slashed D$-$\mathcal{O}(p^4)$ &  $D$-$\mathcal{O}(p^4)$  \\
\colrule
  $m_0$~[MeV]         & $900(6)$      &   $767(6)$     &  $880(22)$      & $908(24)$     \\
  $b_0$~[GeV$^{-1}$]  &$-0.273(6)$    &  $-0.886(5)$   &  $-0.609(19)$   & $-0.744(16)$  \\
  $b_D$~[GeV$^{-1}$]  &$0.0506(17)$   &  $0.0482(17)$  &  $0.225(34)$    & $0.355(20)$   \\
  $b_F$~[GeV$^{-1}$]  &$-0.179(1)$    &  $-0.514(1)$   &  $-0.404(27)$   & $-0.552(28)$  \\
  $b_1$~[GeV$^{-1}$]  & --            &  --            &  $0.550(44)$    & $1.08(6)$   \\
  $b_2$~[GeV$^{-1}$]  & --            &  --            &  $-0.706(99)$   & $0.431(93)$   \\
  $b_3$~[GeV$^{-1}$]  & --            &  --            &  $-0.674(115)$  & $-1.83(15)$   \\
  $b_4$~[GeV$^{-1}$]  & --            &  --            &  $-0.843(81)$   & $-1.57(4)$    \\
  $b_5$~[GeV$^{-2}$]  & --            &  --            &  $-0.555(144)$  & $-0.355(74)$ \\
  $b_6$~[GeV$^{-2}$]  & --            &  --            &  $0.160(95)$    & $-0.423(117)$ \\
  $b_7$~[GeV$^{-2}$]  & --            &  --            &  $1.98(18)$     & $2.79(15)$  \\
  $b_8$~[GeV$^{-2}$]  & --            &  --            &  $0.473(65)$    & $-1.73(6)$ \\
  $d_1$~[GeV$^{-3}$]  & --            &  --            &  $0.0340(143)$  & $0.0157(130)$  \\
  $d_2$~[GeV$^{-3}$]  & --            &  --            &  $0.296(53)$    & $0.445(57)$    \\
  $d_3$~[GeV$^{-3}$]  & --            &  --            &  $0.0431(304)$  & $0.328(18)$ \\
  $d_4$~[GeV$^{-3}$]  & --            &  --            &  $0.234(67)$    & $-0.117(59)$     \\
  $d_5$~[GeV$^{-3}$]  & --            &  --            &  $-0.328(60)$   & $-0.853(77)$    \\
  $d_7$~[GeV$^{-3}$]  & --            &  --            &  $-0.0358(269)$ & $-0.425(39)$    \\
  $d_8$~[GeV$^{-3}$]  & --            &  --            &  $-0.107(32)$   & $-0.557(56)$    \\
  \hline
$\chi^2$/d.o.f. & $11.8$ & $8.6$  & $1.0$ & $1.0$\\
\botrule
\end{tabular} \label{Tab:fitcoef} }
\end{table}

Considering only the octet baryon degrees of freedom, the EOMS BChPT shows a good description of the LQCD data and experimental data with order-by-order improvement. The  $\chi^2/{\rm d.o.f.}$ is about $1.0$ at N$^3$LO, and it indicates that the lattice simulations from these five collaborations are consistent with each other~\footnote{This does not seem to be the case for the LQCD simulations of the g.s. decuplet baryon masses~\cite{Ren:2013oaa}.}, although their setups are totally different. In addition, the values of the fitted LECs turn out to be in a natural range.

Because the octet-decuplet mass splitting, $\delta\sim 0.3$ GeV, is even smaller than the
kaon-pion mass difference, one has to be careful about the virtual decuplet contributions in the BChPT. They
are studied in Ref.~\cite{Ren:2013dzt}  and the corresponding best fit results are given
in Table~\ref{Tab:fitcoef}. The $\chi^2/\mathrm{d.o.f.}$ indicates that the explicit inclusion of the virtual decuplet baryons does not change the description of the LQCD data in any significant way, at least at $\mathcal{O}(p^4)$. On the other hand,  the values of the LECs have changed a lot. This implies that using only the octet baryon mass data, one can not disentangle the virtual decuplet contributions   from  those of the virtual octet baryons and tree-level diagrams~\cite{Ren:2012aj}. However, we notice that the explicit inclusion of the virtual decuplet baryons does
seem to  improve slightly the description of the FVCs, especially for the LQCD data with small $M_{\phi}L$.
\section{Conclusions}

We have studied the lowest-lying octet baryon masses in the EOMS BChPT up to N$^3$LO, taking into account the contributions
of the virtual decuplet baryons.
The unknown low-energy constants are determined by a simultaneous fit to the latest $n_f=2+1$
LQCD simulations from the PACS-CS, LHPC, HSC, QCDSF-UKQCD, and NPLQCD
Collaborations. Finite-volume corrections are calculated self-consistently.

Our studies confirm that the covariant BChPT in the three flavour sector converges as
expected, with clear improvement order
by order. The results also indicate that the LQCD results are consistent
with each other, though their setups are quite different.

We show that in terms of chiral extrapolations, the contributions of the virtual decuplet baryons cannot be easily disentangled from those of the virtual octet baryons and the tree-level contributions, at least at $\mathcal{O}(p^4)$. On the other hand, a slightly better description of finite-volume corrections can be achieved once the virtual decuplet baryons are included.

\section*{Acknowledgments}
This work was partly supported by the National
Natural Science Foundation of China under Grants No. 11005007, No. 11035007, and No. 11175002,  the New Century Excellent Talents in University  Program of Ministry of Education of China under Grant No. NCET-10-0029,  the Fundamental Research Funds for the Central Universities, and
the Research Fund  for the Doctoral Program of Higher Education under Grant No. 20110001110087, the Innovation Foundation of Beihang University for PhD Graduates.

%\begin{thebibliography}{000} %for 3 digits
%\begin{thebibliography}{00}  %for 2 digits

\end{document}